\documentclass[traditabstract]{aa} 
\usepackage{graphicx}
\usepackage{txfonts}
\usepackage{natbib}
\bibpunct{(}{)}{,}{a}{}{,}
\usepackage{subfigure}
\usepackage{nth}
\newcommand{\kms}{\mbox{${\rm km\,s}^{-1}$}}
\newcommand{\ms}{\mbox{${\rm m\,s}^{-1}$}}

\newcommand{\Msolar}{\mbox{${M}_{\sun}$}}
\newcommand{\Rsolar}{\mbox{${R}_{\sun}$}}

\newcommand{\Mast}{\mbox{${M}_{\ast}$}}
\newcommand{\Rast}{\mbox{${R}_{\ast}$}}
\newcommand{\Mjup}{\mbox{${M}_{J}$}}
\newcommand{\rhoast}{\mbox{$\rho_{\ast}$}}

\newcommand{\rhosun}{\mbox{$\rho_{\sun}$}}
\newcommand{\Rjup}{\mbox{${R}_{J}$}}
\newcommand{\rhojup}{\mbox{$\rho_{J}$}}
\newcommand{\teff}{\mbox{$T_{\rm eff}$}}
\newcommand{\logg}{\mbox{$\log g$}}
\newcommand{\vsini}{\mbox{$v \sin i_{\ast}$}}
\newcommand{\mictrb}{\mbox{$\xi_{\rm t}$}}

\newcommand\T{\rule{0pt}{2.2ex}}

\begin{document}

   \title{WASP-117b: a 10-day-period Saturn in an eccentric and misaligned orbit
\thanks{Based on data obtained with WASP-South, CORALIE and EulerCam at the Euler-Swiss telescope, TRAPPIST, 
and HARPS at the ESO 3.6~m telescope (Prog. IDs 087.C-0649, 089.C-0151, 090.C-0540)}$^{,}$
\thanks{The photometric time series and radial velocity data are only available in electronic form
at the CDS via anonymous ftp to cdsarc.u-strasbg.fr (130.79.128.5) or via http://cdsweb.u-strasbg.fr/cgi-bin/qcat?J/A+A/
}}

   \author{M.~Lendl
          \inst{1,2}
          \and
A.H.M.J. Triaud\inst{2,8}
\and
	  D.R. Anderson\inst{3}
\and
A. Collier Cameron\inst{4}
\and
L. Delrez\inst{1}
\and
A.P. Doyle\inst{3}
\and
M. Gillon\inst{1}
\and
C. Hellier\inst{3}
\and
E. Jehin\inst{1}
\and
P.F.L. Maxted\inst{3}
\and
M. Neveu-VanMalle\inst{2,6}
\and
F. Pepe\inst{2}
\and
D. Pollacco\inst{5}
\and
D. Queloz\inst{2,6}
\and
D. S\'egransan\inst{2}
\and
B. Smalley\inst{3}
\and
A.M.S. Smith\inst{3,7}
\and
S. Udry\inst{2} 
\and
V. Van Grootel\inst{1}
\and
R.G. West\inst{5}  
}
\institute{ 
Institut d'Astrophysique et de G\'eophysique, Universit\'e de
Li\`ege, All\'ee du 6 Ao\^ut, 17, Bat. B5C, Li\`ege 1, Belgium
\email{monika.lendl@ulg.ac.be}
\and
Observatoire de Gen\`eve, Universit\'e de Gen\`eve, Chemin des maillettes 51, 1290 Sauverny, Switzerland
\and
Astrophysics Group, Keele University, Staffordshire, ST5 5BG, UK
\and
SUPA, School of Physics and Astronomy, University of St.\ Andrews, North Haugh, Fife, KY16 9SS, UK
\and
Department of Physics, University of Warwick, Gibbet Hill Road, Coventry CV4 7AL, UK
\and
Cavendish Laboratory, J J Thomson Avenue, Cambridge, CB3 0HE, UK
\and
N.~Copernicus Astronomical Centre, Polish Academy of Sciences, Bartycka 18, 00-716 Warsaw, Poland
\and
Kavli Institute for Astrophysics \&\ Space Research, Massachusetts Institute of Technology, Cambridge, MA 02139, USA
}

  \date{}

\abstract{We report the discovery of WASP-117b, the first planet with a period beyond 10~days 
found by the WASP survey. The planet has a mass of $M_p= 0.2755 \pm 0.0089 \, \Mjup$, a radius of 
$R_p= 1.021_{-0.065}^{+0.076}\, \Rjup$ and is in an eccentric ($ e= 0.302 \pm 0.023 $), $ 10.02165 \pm 0.00055 $~d 
orbit around a main-sequence F9 star. The host star's brightness (V=10.15~mag) makes WASP-117 
a good target for follow-up observations, and with a periastron planetary equilibrium temperature of 
$T_{eq}= 1225_{-39}^{+36}$~K and a low planetary mean density ($\rho_p= 0.259_{-0.048}^{+0.054} \, \rhojup$) it 
is one of the best targets for transmission spectroscopy among planets with periods around 10
days. From a measurement of the Rossiter-McLaughlin effect, we infer a projected angle between the 
planetary orbit and stellar spin axes of $\beta = -44 \pm 11$~deg, and we further derive an orbital 
obliquity of $\psi = 69.6 ^{+4.7}_{-4.1}$~deg. 
Owing to the large orbital separation, tidal forces causing orbital circularization and 
realignment of the planetary orbit with the stellar plane are weak, having had little impact on the
planetary orbit over the system lifetime. WASP-117b joins a small sample of transiting giant planets with
well characterized orbits at periods above 8 days.

 }

   \keywords{planetary systems -- stars: individual: WASP-117 -- techniques: spectroscopic -- techniques: photometric}

   \maketitle
%

\section{Introduction}

Transiting planets play a fundamental part in the study of exoplanets and planetary systems
as their radii and absolute masses can be measured.
For bright transiting systems, several avenues of follow-up observations can be pursued, giving access to
the planetary transmission and emission spectrum, and the systems obliquity (via the sky-projected angle between the 
stellar spin and planetary orbit; see e.g. \citet{Winn11} for a summary).
Ground-based transit surveys such as WASP \citep{Pollacco06}, HAT \citep{Bakos04}, and KELT \citep{Pepper07} have 
been key in discovering the population of \textit{hot Jupiters}, giant planets with periods of only
a few days.

There are two main theories of the inward migration of gas giants to create these \textit{hot Jupiters}.
Disk-driven migration (\citealt{Goldreich80,Lin86}, see \citealt{Baruteau13} for a summary) relies on the 
interaction of a gas giant with the protoplanetary disk to cause inward migration, producing giant planets 
on short-period circular orbits. The obliquities produced from disc migration are inherited from the 
inclination of the protoplanetary disc. While this mechanism favors low orbital obliquities, stellar binary 
companions may cause the protoplanetary disc to tilt with respect to the stellar equator, leading to the creation of misaligned planets
\citep{Batygin12,Lai14}. Dynamical migration processes, such as planet-planet scattering (\citealt{Rasio96a,Weidenschilling96}, 
see \citealt{Davies13} for a summary) or migration through Lidov-Kozai cycles (\citealt{Lidov62,Kozai62,Eggleton01,Wu03}),
require the planet to be placed on a highly eccentric orbit, that is subsequently circularized by tidal 
interactions. This migration pathway can produce large orbital obliquities, 
such as those observed for several giant planets \citep[e.g.][]{Hebrard08,Triaud10,Winn10}. 
It has been suggested that obliquities are damped by tidal interactions between planet
and host star, the efficiency of this process together with the system age reproducing the observed 
obliquity distribution \citep{Winn10,Triaud11,Albrecht12}.

We present WASP-117b, the planet with the longest period and one of the lowest masses 
(surpassed only by WASP-29b, \citealt{Hellier10}) found by the WASP survey to date. 
This Saturn-mass planet is in an eccentric and misaligned 10.02 day-period orbit around an F9 star, and 
is one of the few transiting gas giants known with periods in the range of 8 -- 30~days. 

\section{Observations}
\label{sec:obs}

\subsection{WASP photometry}
WASP-117 (GSC08055-00876) was observed with the WASP-South facility throughout the 2010 and 2011 seasons,
leading to the collection of 17933 photometric measurements. The photometric reduction and target selection processes are 
described in more detail in \citet{Pollacco06} and \citet{Cameron07}. A periodic dimming was detected using the algorithms described
in \citet{Cameron06}, leading to the selection of the target for spectroscopic and photometric follow-up. The phase-folded
discovery lightcurve is shown in Figure \ref{fig:RV}.

\subsection{Spectroscopic follow-up}

\begin{figure}
 \centering
 \includegraphics[width=0.95\linewidth]{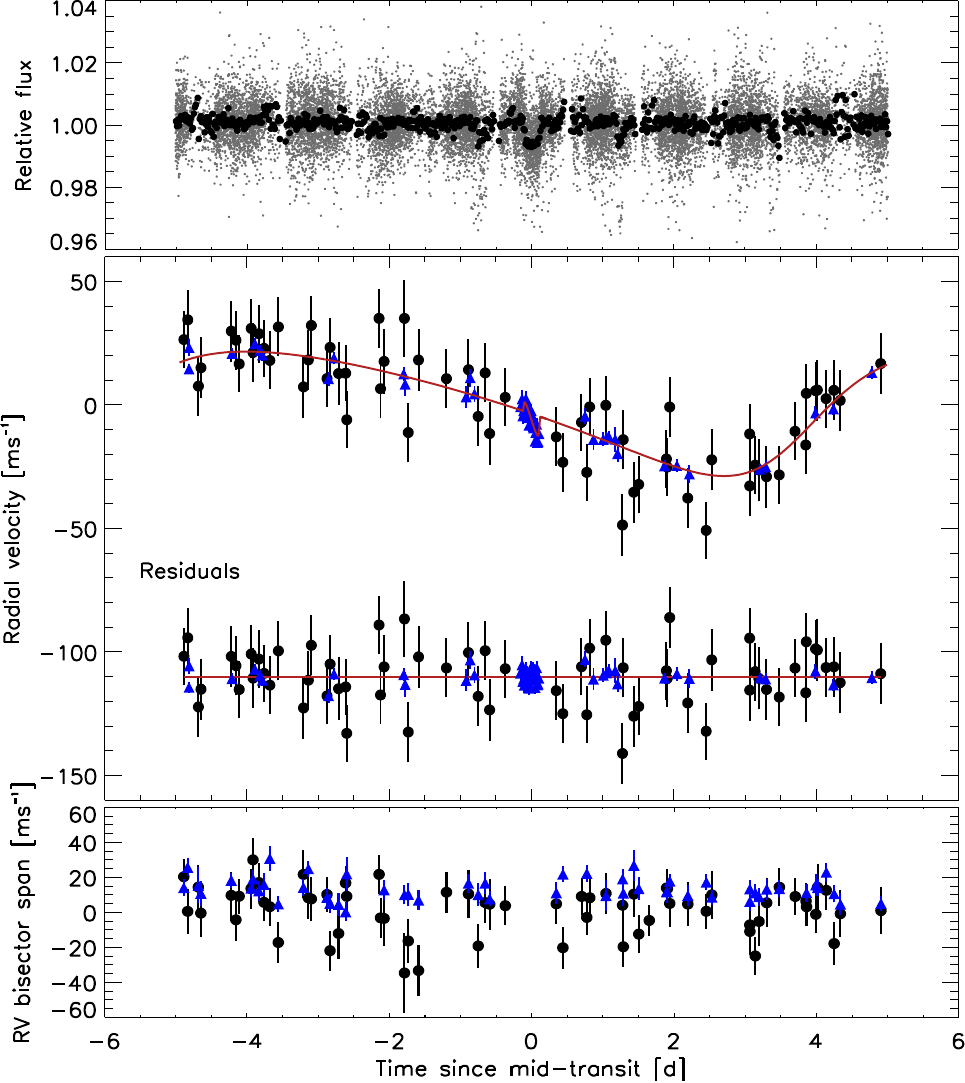}
 \caption{\label{fig:RV}\textit{Top:} The WASP lightcurve of WASP-117 folded on the planetary period. All data are shown
in gray, and the phase-folded data, binned into 20~min intervals, are shown in black. \textit{Middle:} The CORALIE (black circles) and 
HARPS (blue triangles) radial velocities folded on the planetary period together with the best-fitting model (red solid line) and residuals.
\textit{Bottom:} The bisector spans of the above RV data.}
\end{figure}

We used CORALIE at the 1.2~m Euler-Swiss and HARPS at the ESO 3.6~m telescopes for spectroscopic measurements of WASP-117,
obtaining 70 (CORALIE) and 53 (HARPS) target spectra. 
Radial velocities (RVs) were determined using the weighted cross-correlation method \citep{Baranne96,Pepe02}. 
We detected a RV variation with a period of \textasciitilde10~days compatible 
with that of a planet, and verified its independence of the measured bisector span as suggested by \citet{Queloz00}.
The RV data are shown in Figure \ref{fig:RV}. Once the planetary status was confirmed, HARPS was used to observe the system 
throughout a transit, aiming to measure the projected spin-orbit angle via the Rossiter-McLaughlin 
effect \citep{Rossiter24,McLaughlin24}. These data are shown in Figure \ref{fig:TRA}.
\subsection{Photometric follow-up}

\begin{figure}
 \centering
 \includegraphics[width=0.95\linewidth]{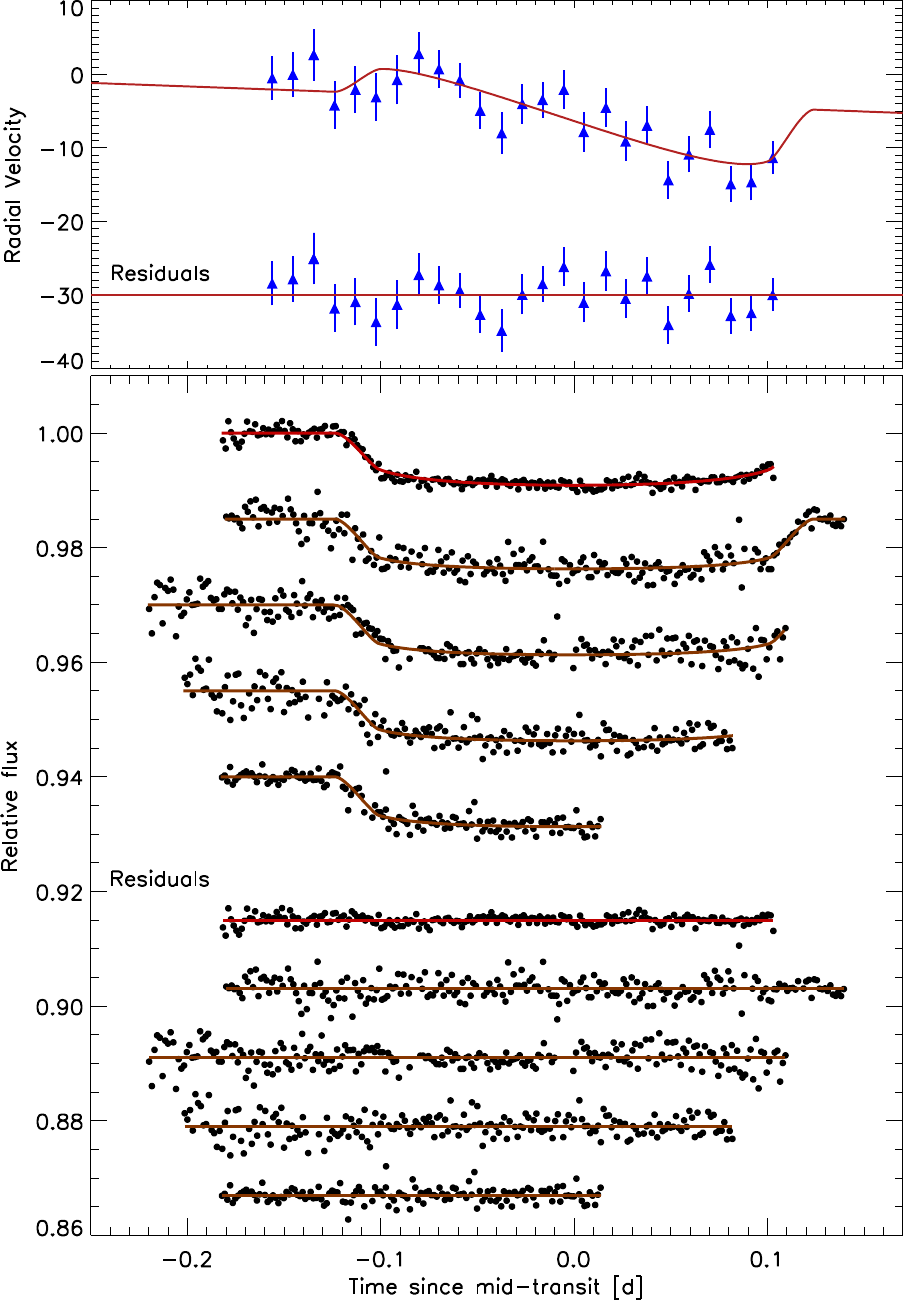}
 \caption{\label{fig:TRA}\textit{Top:} HARPS RVs obtained on 29 Aug 2013, during a transit of WASP-117b 
together with the best-fitting model and residuals. \textit{Bottom:} The transit lightcurves observed with (from top to bottom)
EulerCam on 29 Aug 2013, and TRAPPIST on 19 Aug 2013, 29 Aug 2013, 08 Sep 2013, and 28 Sep 2013 (all UT). 
The data are offset vertically for clarity. The lightcurves are corrected for their photometric baseline models, the
best-fitting transit models are superimposed and the residuals shown below.}
\end{figure}

We obtained a total of five high-precision transit lightcurves. Four were observed with the 60~cm TRAPPIST telescope 
\citep{Gillon11a,Jehin11} using a \textit{z'-Gunn} filter, and one was obtained with EulerCam at the 1.2~m Euler-Swiss 
telescope using an \textit{r'-Gunn} filter. All photometric data were extracted using relative aperture photometry, while
carefully selecting optimal extraction apertures and reference stars. More details on EulerCam and the reduction of EulerCam data 
can be found in \citet{Lendl12}. The resulting lightcurves are shown in Figure \ref{fig:TRA}.

\section{Determination of system parameters}
\label{sec:det}
\subsection{Stellar parameters}
\label{sec:stel}

The spectral analysis was performed based on the HARPS data. We used the standard pipeline reduction products and 
co-added them to produce a single spectrum (S/N=240). We followed the procedures outlined in \citet{Doyle13} 
to obtain the stellar parameters given in Table \ref{tab:stel}. For the vsini determination we have assumed a 
macroturbulence of $4.28 \pm 0.57$~{\kms} obtained using the asteroseismic-based calibration of \citet{Doyle14}.
The lithium abundance suggests an age between 2 and 5 Gyr \citep{Sestito05}.

To deduce the stellar mass and estimate the system age, we performed a stellar evolution modeling based on the CLES code 
(\citealt{Scuflaire08}; see \citealt{Delrez14} for details). Using as input the stellar mean density deduced from the global analysis described
in Section \ref{sec:glo}, and the effective temperature and metallicity from spectroscopy, we obtained a stellar mass of 
1.03 $\pm$ 0.10 $M_{\odot}$ and an age of 4.6 $\pm$ 2.0 Gyr. This value agrees with the age inferred from the lithium abundance
and places WASP-117 on the main sequence. The error budget is dominated by uncertainties affecting the stellar 
internal physics, especially for the initial helium abundance of WASP-117. 

\begin{table}[h]
\centering   
\caption{\label{tab:stel}Stellar parameters of WASP-117 including those obtained from the spectroscopic analysis 
described in Section \ref{sec:stel}.
\newline Note: The age is that obtained from stellar evolution modeling, and the spectral type was estimated from {\teff}
using the table in \citet{Gray08}. The {[Fe/H]} abundance is relative to the Solar value given in \citet{Asplund09}.}
\begin{tabular}{llll} \hline \hline
RA   & $02^{\mathrm{h}}27^{\mathrm{m}}06.09^{\mathrm{s}}$ & V mag      &  10.15   \T \\
DEC & $-50^{\circ}17'04.3''$  & & \\ \hline
\teff [K] &  6040 $\pm$ 90~K & \logg      &  4.28 $\pm$ 0.16 \T\\
\mictrb [\kms] &  0.96 $\pm$ 0.14 & \vsini [\kms] &  1.55 $\pm$ 0.44\\
{[Fe/H]} &  $-0.11 \pm$ 0.14 & log A(Li)  &  2.21 $\pm$ 0.05 \\
Age [Gyr] &  4.6 $\pm$ 2.0  & Spectral Type   &  F9 \\
$\log R^{\prime}_{\rm HK}$ & $-4.95 \pm 0.05 $ & $P_{rot}$ [d]  & $17.1 \pm 2.6$ \\
\hline
\end{tabular}
\end{table}

We checked the WASP lightcurves for periodic modulation such as caused by the interplay of stellar activity and 
rotation using the method described in \citet{Maxted11}. No variations exceeding 95\% significance 
were found above 1.3~mmag. The low activity level of the star is confirmed by a low activity index,
$\log R^{\prime}_{\rm HK} = -4.95 \pm 0.05$. Using the relation of \citet{Mamajek08} that connects chromospheric
activity and stellar rotation, we estimate a stellar rotation period of $P_{rot} = 17.1 \pm 2.6 $ days. 

\label{sec:evo}

\subsection{Global analysis}
\label{sec:glo}

\subsubsection{MCMC}

We applied a Markov chain Monte Carlo (MCMC) approach to derive the system parameters. Details on MCMCs in astrophysical
contexts can be found e.g. in \citet{Tegmark04}. To do so, we carried out a simultaneous analysis of
data from the photometric and spectroscopic follow-up observations. We made use of the MCMC implementation described in detail in 
\citet{Gillon12a}, with the fitted (``jump'') parameters listed in Table \ref{tab:par}. The transit lightcurves were 
modeled using the prescription of \citet{Mandel02} as well as a quadratic model to describe stellar limb-darkening. 
A Keplerian together with the prescription of the Rossiter-McLaughlin effect by \citet{Gimenez06} was used for for the RV data.
The convective blueshift effect \citep{Shporer11} was not included in the model as its amplitude (\textasciitilde$\, 1\, {\ms}$) 
is well below our 1~$\sigma$ errors.
We used the results obtained from the stellar spectral analysis to impose normal prior distributions on the stellar parameters {\teff}, 
{\vsini}, and {[Fe/H]}, with centers and widths corresponding to the quoted measurements and errors, respectively. Similarly, a normal 
prior was imposed on the limb-darkening coefficients using values interpolated from the tables by \citet{Claret11}. For the other 
parameters uniform prior distributions were used.

\subsubsection{Photometry}
We estimated the correlated (red) noise present in the data via the $\beta_r$ factor \citep{Winn08,Gillon10a}, and measured excess white 
noise, $\beta_w$, by comparing the calculated error bars to the final lightcurve RMS. The photometric error bars were then 
rescaled by a factor $CF = \beta_r \times \beta_w$ for the final analysis. 
The minimal photometric baseline variation necessary to correct for photometric trends was assumed to be a \nth{2}-order time polynomial, and 
an offset at the meridian flip for the TRAPPIST lightcurves. For the TRAPPIST data observed on 28 September 2013, a 
significant improvement was found by additionally including a \nth{1}-order coordinate dependence. 

\subsubsection{Radial velocities}
To compensate for RV jitter affecting the CORALIE data, we added 10.4~{\ms} quadratically to the CORALIE error bars. The HARPS data did 
not require any additional jitter. The RV zero points were fitted via minimization at each MCMC step.
The inferences drawn from the Rossiter-McLaughlin effect are highly sensitive to any potential RV offsets around the time of transit, 
as can be caused by stellar activity. Throughout the HARPS RV sequence obtained during the transit of 29 Aug 2013, we obtained simultaneous 
photometric observations from EulerCam and TRAPPIST that serve to detect spot-crossing events as well as add information on the precise 
timing of the transit contact points. No signatures of spot-crossing events are visible in the photometry, in line with the low 
$\log R^{\prime}_{\rm HK}$ activity index observed. 
To test for any potential RV offset that may still be present, we 
treated the HARPS data obtained during the night of transit (25 in-, and 3 out-of-transit points) as well as the nights before (1 point) 
and after (2 points) the transit as independent sequences, allowing for independent RV zero points. The resulting HARPS RV zero points are 
$\gamma_{\rm HAR,1} = -16.03398_{-0.0011}^{+0.0018}\,\kms$ (HARPS, in and near transit), and 
$\gamma_{\rm HAR,2} = -16.03268_{-0.00011}^{+0.00013}\, \kms$ HARPS (other). 
The two zero points obtained from the HARPS data are in good agreement with each other. Thus we further allow only one RV zero point per
instrument, finding $\gamma_{\rm HAR} = -16.03292_{-0.00095}^{+0.00060} \, \kms$ and $\gamma_{\rm COR} = -16.044910 \pm 4.7 \times 10^{-5}\, \kms$.

We verified the independence of the Keplerian solution from the in-transit points by carrying out a global analysis excluding 
the in-transit RVs. The resulting parameters are less than $0.1 \sigma$ from those found from the analysis of the entire dataset (see Section 
\ref{sec:spar}). Similarly, excluding the CORALIE points from the analysis did not produce any significant changes.

\begin{figure*}
 \centering
 \includegraphics[width=\linewidth]{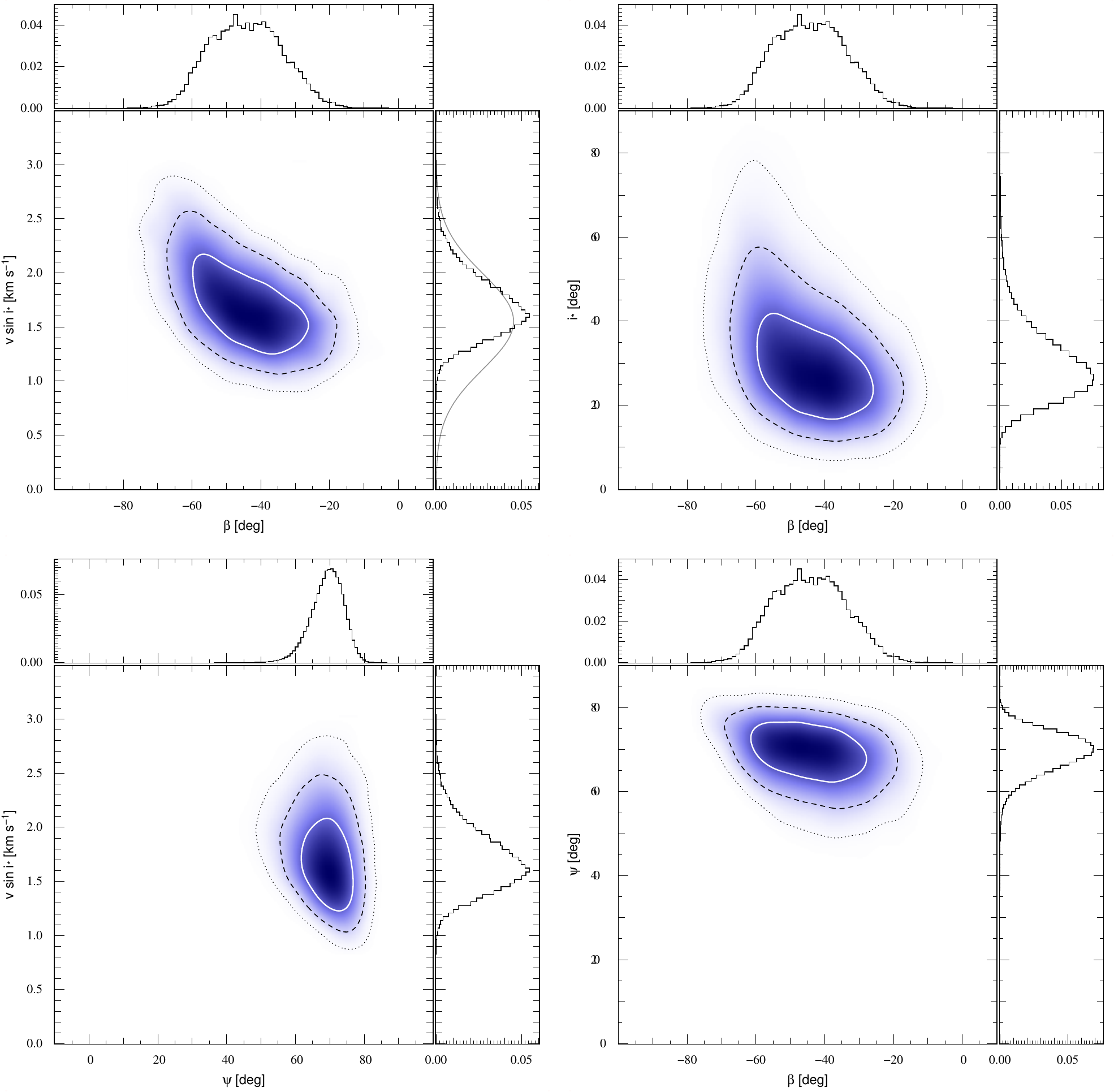}
 \caption{\label{fig:bpsi}Posterior probability distributions of the parameters $\beta$, $\vsini$, $i_\ast$, and $\psi$. 
It is noteworthy that the low projected obliquities $\beta$ coincide with low values of the stellar inclination $i_\ast$ 
(i.e., a star seen nearly pole-on) and hence a large true orbital obliquity.}
\end{figure*}   

\subsubsection{System parameters}
\label{sec:spar}

The system parameters given in Table \ref{tab:par} were found by running two MCMC chains of 100000 points each, whose convergence was checked with the 
Gelman \& Rubin test \citep{Gelman92}. As we have knowledge of the parameters $e$ and $\omega$ from radial velocities, we used the transit 
lightcurves to obtain a measurement of {\rhoast} \citep{Kipping10}. We then inferred the stellar parameters {\Mast} and {\Rast} based on the 
stellar {\teff}, [Fe/H] and {\rhoast} using the calibration of \citet{Enoch10} that is based on a fit to a sample of main-sequence eclipsing 
binaries. The stellar parameters obtained from this calibration are in good agreement with those found by stellar evolution modeling.

We find a projected obliquity of $\beta = -44 \pm 11 $~deg that is moderately correlated with the stellar $v\sin{i_\ast}$, 
as shown in Figure \ref{fig:bpsi}. The true obliquity, $\psi$, can be computed if we have information on $\beta$, $i_p$, and $i_\ast$ 
\citep{Fabrycky09}. Combining our posterior distributions of {\Rast} and {\vsini} with the estimate of $\mathrm{P_{rot}}$ obtained in 
Section \ref{sec:stel}, we calculate $ i_\ast = 28.6^{+6.1}_{-8.5}$~deg. This value, together with the posterior distributions 
of $\beta$ and $i_p$, yields $\psi = 69.6 ^{+4.7}_{-4.1}$~deg. As shown in Figure \ref{fig:bpsi}, $\beta$ and $i_\ast$ are 
correlated such that low values for $\beta$ correspond to low values of $i_\ast$, i.e. near pole-on stellar orientations.
WASP-117b is clearly misaligned.

\begin{table}
\centering                        
\caption{\label{tab:par}Planetary and stellar parameters for WASP-117 from a global MCMC analysis. 
\newline \tablefoottext{a}{Equilibrium temperature, assuming A=0 and F=1 \citep{Seager05}.}
\tablefoottext{b}{Orbital separation averaged over time.}}
\begin{tabular}{p{5.3cm}p{2.8cm}}       
\hline\hline 
 \multicolumn{2}{l}{Jump parameters} \T  \\
\hline
 Transit depth, $ \Delta F$ \T & $ 0.00803_{-0.00048}^{+0.00055} $ \\
 $ b' = a*\cos(i_{p})$ $[R_{\ast}]$    & $ 0.262_{-0.089}^{+0.074} $ \\
 Transit duration, $ T_{14}$ [d] &  $ 0.2475_{-0.0029}^{+0.0033} $ \\
 Mid-transit time, [HJD] - 2450000 & $ 6533.82326_{-0.00090}^{+0.00095} $ \\
 Period, $ P$ [d] & $ 10.02165 \pm 0.00055  $ \\
 $ K_2=K\sqrt{1-e^2}P^{1/3}$~[$\mathrm{ms^{-1}d^{1/3}}$] & $ 51.7 \pm 1.4 $ \\
 $ \sqrt{e}\cos\omega $ & $ -0.258 \pm 0.016  $ \\
 $ \sqrt{e}\sin\omega $ & $ -0.485_{-0.026}^{+0.029} $ \\
 $ \sqrt{\vsini}\cos{\beta}$ & $ 0.91 \pm 0.14 $ \\
 $ \sqrt{\vsini}\sin{\beta}$ & $ -0.89 \pm 0.23 $ \\
 Stellar eff. temperature, $ \teff_{\ast} $ [K] & $ 6038 \pm 88  $ \\
 Stellar metallicity, $ \mathrm{[Fe/H]}_{\ast} $ & $ -0.11 \pm 0.14 $ \\
 $ c_{1,\rm r'}=2u_{1,\rm r'}+u_{2,\rm r'} $   & $ 0.975 \pm 0.031 $ \\
 $ c_{2,\rm r'}=u_{1,\rm r'}-2u_{2,\rm r'} $   & $ -0.263 \pm 0.018 $ \\
 $ c_{1,\rm z'}=2u_{1,\rm z'}+u_{2,\rm z'} $   & $ 0.705 \pm 0.022 $ \\
 $ c_{2,\rm z'}=u_{1,\rm z'}-2u_{2,\rm z'} $   & $ -0.352 \pm 0.012 $ \\
\hline 
 \multicolumn{2}{l}{Deduced parameters} \T \\
\hline
 RV amplitude, $ K $ [{\ms}] \T    & $ 25.16 \pm 0.69 $ \\
 RV zero point (HARPS), $\gamma_{\rm HAR} $  [{\kms}] & $ -16.03292_{-0.00095}^{+0.00060}$ \\
 RV zero point (CORALIE), $\gamma_{\rm COR} $  [{\kms}] & $ -16.044910 \pm 4.7 \times 10^{-5} $ \\
 Planetary radius, $ R_{p} $ [{\Rjup}] & $ 1.021_{-0.065}^{+0.076}  $ \\
 Planetary mass, $ M_{p} $ [{\Mjup}] & $ 0.2755 \pm 0.0090 $ \\
 Planetary mean density, $ \rho_{p} $ [{\rhojup}]  & $ 0.259_{-0.048}^{+0.054} $\\
 Planetary grav. acceleration, $ \logg_{p} $ [cgs] & $ 2.817_{-0.058}^{+0.054} $ \\
 Planetary eq. temperature, $ T_{eq} $ [K]\tablefootmark{a,b} & $ 1001_{-32}^{+29} $ \\
 Planetary eq. temperature at periastron, $ T_{eq} $ [K]\tablefootmark{a} & $ 1225_{-39}^{+36} $ \\
 Planetary eq. temperature at apoastron, $ T_{eq} $ [K]\tablefootmark{a} & $ 897_{-29}^{+26} $ \\
 Orbital semi-major axis, $ a $ [au] & $ 0.09459_{-0.00079}^{+0.00084} $ \\
 $ a/R_{\ast} $  & $ 17.39 \pm 0.81 $ \\
 Eccentricity, $ e $   & $ 0.302 \pm 0.023 $ \\
 Argument of periastron, $ \omega $ [deg] & $ 242.0_{-2.7}^{+2.3} $ \\
 Inclination, $ i_{p} $ [deg] & $ 89.14 \pm 0.30 $\\
 Transit impact parameter, $ b_{tr} $  &  $ 0.32_{-0.11}^{+0.09 } $ \\
 Proj. orbital obliquity [deg], $\beta$ & $ -44 \pm 11 $ \\
 Orbital obliquity [deg], $\psi$ & $ 69.6 ^{+4.7}_{-4.1} $ \\
 Stellar mass, $ M_{\ast} $ [{\Msolar}]    & $ 1.126 \pm 0.029 $ \\
 Stellar radius, $ R_{\ast} $ [{\Rsolar}]    & $ 1.170_{-0.059}^{+0.067} $ \\
 Stellar mean density, $ \rho_{\ast} $ [{\rhosun}] & $ 0.70\pm 0.10 $\\
 Stellar $\vsini$ [{\kms}] & $ 1.67_{-0.24}^{+0.31} $ \\ 
 Stellar inclination, $i_\ast$ [deg] &  $ 28.6^{+6.1}_{-8.5} $  \\
 Limb-darkening coefficient, $ u_{1,\rm r'} $    & $ 0.337 \pm 0.016 $ \\
 Limb-darkening coefficient, $ u_{2,\rm r'} $    & $ 0.2997 \pm 0.0058 $ \\
 Limb-darkening coefficient, $ u_{1,\rm z'} $    & $ 0.211 \pm 0.011 $ \\
 Limb-darkening coefficient, $ u_{2,\rm z'} $    & $ 0.2818 \pm 0.0033 $ \\ 
\hline
\end{tabular}
\end{table}

\section{Discussion}
\label{sec:dis}

\begin{figure*}
 \includegraphics[width=\linewidth]{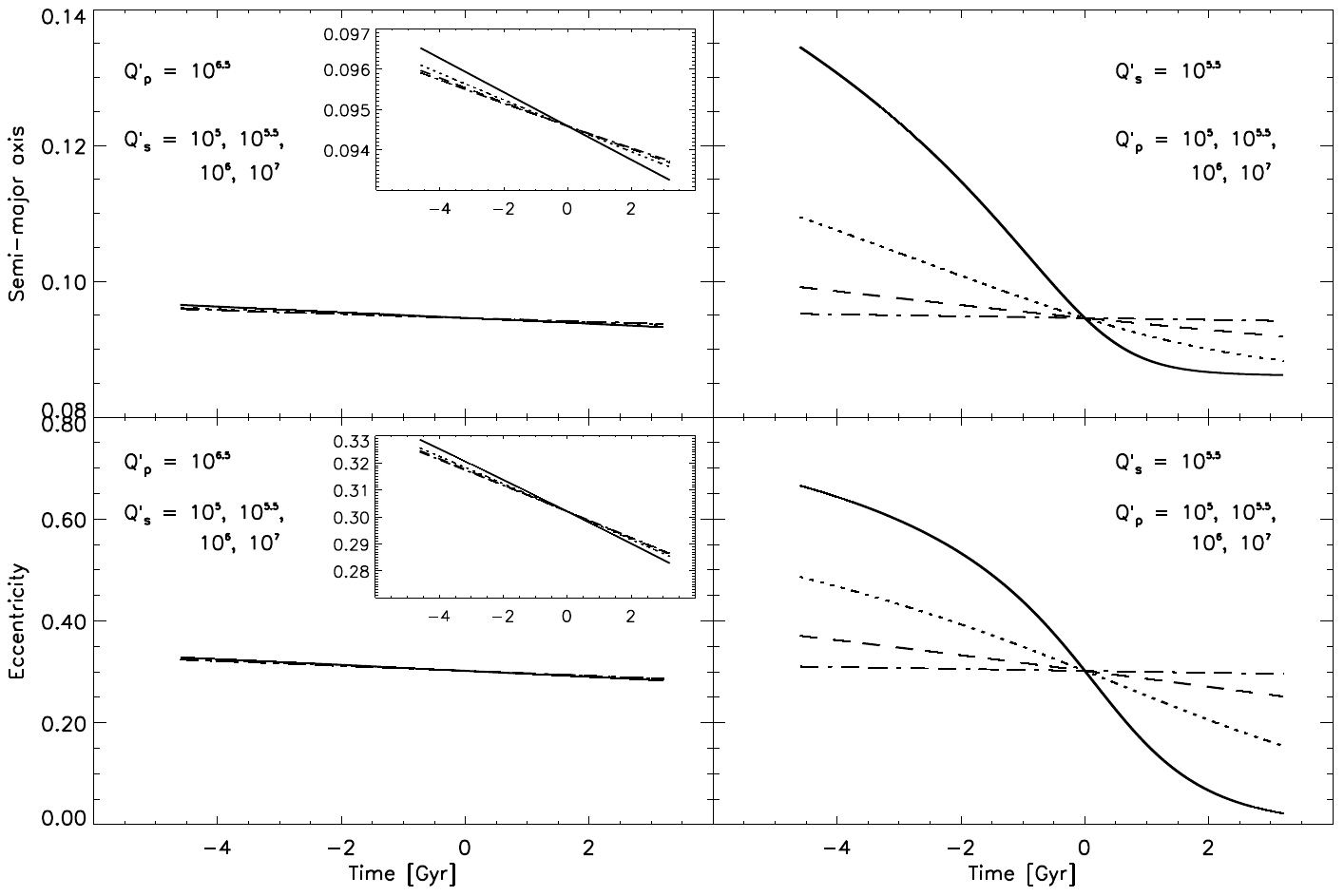}
 \caption{\label{fig:evol}Tidal evolution of the orbital separation (top) and eccentricity (bottom) of WASP-117~b, calculated both forwards and backwards in time.
 \textit{Left:} $Q'_p$ was set to a constant value of $Q'_p=10^{6.5}$, while $Q'_s$ values of $10^5$ (solid line), $10^{5.5}$ (dotted line), $10^6$ 
(dashed line), and $10^7$ (dash-dotted line) were used. The subplots show a zoom to visualize the variations. \textit{Right:} $Q'_s$ was set to a constant value of $Q'_s=10^{5.5}$, while $Q'_p$ values 
of $10^5$ (solid line), $10^{5.5}$ (dotted line), $10^6$ (dashed line), and $10^7$ (dash-dotted line) were used.}
\end{figure*}

With a period of \textasciitilde10~days, WASP-117b resides in a less densely populated parameter space, well outside the pile-up of giant planets
at periods below 5~days. In terms of mean density, this Saturn-mass planet is no outlier with respect to 
other transiting planets of the same mass range. The orbital eccentricity is well constrained at $e= 0.302 \pm 0.023$. We measure a 
projected spin-orbit angle of $\beta = -44 \pm 11$~deg and infer an obliquity of $\psi = 69.6 ^{+4.7}_{-4.1}$~deg.

Owing to its large orbital separation, tidal interactions between star and planet are weaker than for close-in systems. This means that the 
dampening of eccentricities and orbital decay via tidal interactions is greatly reduced. We integrated Equations (1) and (2) of \citet{Jackson08} 
both backwards and forwards in time for WASP-117, testing $Q'_p$ and $Q'_s$ values between $10^5$ and $10^7$. We assumed a system age
of 4.6~Gyr as well as a main-sequence lifetime of 7.8~Gyr as inferred from the stellar evolution modeling (see Section \ref{sec:stel}).
The resulting tidal evolutions are shown in Figure \ref{fig:evol}.
For all tested $Q'_s$ values, the overall changes in $a$ and $e$ are below 0.004~au and 0.05, respectively, over the systems main-sequence 
lifetime. Varying $Q'_p$ substantially from the suggested value of $Q'_p=10^{6.5}$ \citep{Jackson08} allows for 
some evolution in the orbital eccentricity, but not complete circularization. For the most extreme case studied, $Q'_p = 10^{5}$,
an initial eccentricity of 0.67 evolves down to 0.02 over the systems lifetime, together with the orbital separation evolving from 
0.135~au to 0.086~au. 

Similarly, tidal interactions have been invoked to realign initially misaligned planetary systems \citep{Winn10b}.
To estimate the extent to which this process has impacted the WASP-117 system, we use Equation (4) of \citet{Albrecht12} to calculate the 
efficiency of spin-orbit realignment, assuming a mass of $0.009 \Mast$ in the stellar convective envelope \citep{Pinsonneault01}. 
The resulting value $\tau_{Mcz} = 4.6\times 10^{12}$~yr, indicates that the orbital obliquity has been essentially unchanged by tidal interactions.

The weak tidal interactions explain why this is one of the few inclined systems amongst stars colder than 6250 K \citep{Winn10b}. 
It worth remarking that the other inclined planets orbiting cold stars are also eccentric (e.g. HD80606b, \citealp{Hebrard08};
WASP-8Ab, \citealp{Queloz10}; and HAT-P-11b, \citealp{Winn10}).
The independence of orbital eccentricity and obliquity from stellar tides make WASP-117b a valuable object for the understanding 
of planetary migration, and its connection to the properties of close-in planets. It appears that the planet has not undergone an episode of 
high eccentricity during its migration, as tidal interactions are not strong enough to efficiently circulate a highly eccentric orbit at these 
large orbital separations. In order to judge the influence of dynamical interactions on the evolution of this system, we encourage 
follow-up observations aimed at detecting further stellar or planetary companions.

Among the planets outside of the short-period pileup, WASP-117b is one of the most favorable targets for atmospheric characterization, 
in particular through transmission spectroscopy. This is due to both, a calm, bright (V=10.15~mag) host star, and an extended planetary atmosphere. 

\section{Conclusions}
\label{sec:con}

The 10-day-period transiting Saturn-mass planet WASP-117b represents the longest-period and one of the lowest-mass planets found by the 
WASP survey to date. The planetary orbit is eccentric and is inclined with respect to the stellar equator. Tidal
interactions between planet and host star are unlikely to have substantially modified either of these parameters over the systems lifetime,
making WASP-117b an important piece in the puzzle of understanding the connection between migration and the currently observed
orbital properties of close-in planets.

\begin{acknowledgements}
WASP-South is hosted by the South African Astronomical Observatory 
and we are grateful for their ongoing support and assistance. 
Funding for WASP comes from consortium universities
and from the UK's Science and Technology Facilities Council.
TRAPPIST is funded by the Belgian Fund for Scientific  
Research (Fond National de la Recherche Scientifique, FNRS) under the  
grant FRFC 2.5.594.09.F, with the participation of the Swiss National  
Science Fundation (SNF). M. Gillon and E. Jehin are FNRS Research  
Associates. This work was supported by the European Research 
Council through the European Union's Seventh Framework Programme 
(FP7/2007-2013)/ERC grant agreement number 336480. L. Delrez acknowledges 
the support of the F.R.I.A. fund of the FNRS. M. Gillon and E. Jehin are 
FNRS Research Associates. A.H.M.J. Triaud received funding from the 
Swiss National Science Foundation under grant number P300P2-147773.
\end{acknowledgements}

\bibliographystyle{aa}
\bibliography{bbl}

\end{document}